\documentclass[conference,9pt]{IEEEtran}
\usepackage{float}
\usepackage{comment}
\usepackage[table]{xcolor}
\usepackage{cite}      
\usepackage{graphicx}  
\usepackage{psfrag}    
\usepackage{placeins}
\usepackage{lineno}
\usepackage{url}
\usepackage{subcaption}
\usepackage{caption}
\usepackage{times}
\usepackage{textcomp}
\usepackage{amsmath} 
\usepackage{amsfonts,amsthm,amssymb,mathtools}
\usepackage{balance}
\usepackage[normalem]{ulem}
\usepackage{xcolor}

\usepackage{cuted}
\usepackage{amsmath,bm}
\usepackage{hyperref}
\usepackage{cleveref}
\begin{document}
\title{Estimation of Electrical Characteristics of Complex Walls Using Deep Neural Networks}

    
\author{\IEEEauthorblockN{Kainat~Yasmeen and Shobha~Sundar~Ram\\
Indraprastha Institute of Information Technology Delhi, New Delhi 110020 India\\
Email: \{kainaty, shobha\}@iiitd.ac.in}}

\maketitle

\begin{abstract}
Electromagnetic wave propagation through complex inhomogeneous walls introduces significant distortions to through-wall radar signatures. Estimation of wall thickness, dielectric, and conductivity profiles may enable wall effects to be deconvolved from target scattering. We propose to use deep neural networks (DNNs) to estimate wall characteristics from broadband scattered electric fields on the same side of the wall as the transmitter. We demonstrate that both single deep artificial and convolutional neural networks and dual networks involving generative adversarial networks are capable of performing the highly nonlinear regression operation of electromagnetic inverse scattering for wall characterization. These networks are trained with simulation data generated from full wave solvers and validated on both simulated and real wall data with approximately 95\% accuracy. 
\end{abstract}
\providecommand{\keywords}[1]{\textbf{\textit{Keywords--}}#1}

\section{Introduction}
\label{sec:Introduction}
Through-wall radar signals undergo complex propagation phenomenology, including reflection, refraction, attenuation, ringing, and multipath, which produces ghost targets and other forms of significant distortions in the radar signatures \cite{ram2009simulation,ram2016through}. In order to remove these distortions, we must be able to deconvolve the wall effects from the target-scattered signal. However, walls are considerably diverse, ranging from single-layer homogeneous dielectric indoor walls to inhomogeneous/multi-layered outer walls. Wall materials vary from brick, stone, mud, wood, and cement to cinder blocks with air gaps. Various strategies have been investigated to deal with radar signatures distorted by walls. The authors in \cite{solimene2009three} remove the front face reflections from homogeneous single-layer dielectric walls from the through-wall target scattered signals. Later, sparsity-based deep learning techniques were exploited to eliminate wall effects with and without the requirement for prior knowledge of wall characteristics \cite{liu2019clutter,vishwakarma2020mitigation,ram2021sparsity}. In \cite{setlur2011multipath}, the multipath introduced by through-wall propagation is exploited to improve the radar detection performance. Radar hardware parameters are adapted to abolish wall effects from radar signatures in \cite{guo2018multipath,zheng2018multilevel}. In this work, we focus on the specific problem of estimating a wall's electrical characteristics based on the scattered electric field. This estimate can subsequently be used to remove the wall effects from the signatures. The problem falls within the broader domain of electromagnetic inverse scattering (EIS). 

EIS is the method of estimating the electrical characteristics of an unknown object in a region of interest based on its scattered electric field. Over the past decade, EIS has been researched due to its wide range of civilian and military applications, such as radar imaging, biomedical imaging, non-destructive testing, and material characterization \cite{chen2020review, liu2018generative, li2018deepnis}. 
 Historically, deterministic methods like the Born approximation technique \cite{habashy1993beyond}, or backpropagation \cite{chen2018computational} have been used to tackle EIS problems. However, it is well known that the accuracy of these methods is low, especially when the region of interest contains scatterers with high dielectric constants. Iterative microwave imaging approaches have also been investigated, including the Born/Rytov iterative methods \cite{wang1989iterative}, the distorted Born iterative method \cite{chew1990reconstruction}, the contrast source-type inversion method \cite{song2005through} and the subspace optimization method \cite{chen2009subspace}. These methods take a lot of time and memory and are unsuitable for real-time reconstruction. Further, all of these methods have been restricted to the estimation of the dielectric profile of the scatterer while assuming that it is inherently lossless. 
 
 An alternative approach that has been explored is to use artificial neural network-based methodologies to extract information about the geometric and electromagnetic properties of scatterers \cite{caorsi1999electromagnetic, rekanos2002neural}. These methods typically estimate a few wall parameters - such as their width and relative permittivities - for homogeneous profiles only. Recently, deep learning methods have been explored for solving highly non-linear and ill-posed problems in diverse electromagnetics problems such as remote sensing, microwave imaging, ground penetration radar, and biomedical imaging \cite{chen2020review}, radar classification \cite{alnujaim2021synthesis}, and through-wall radar imaging \cite{cicchetti2021numerical,zheng2021human,li2020human}. In this paper, we propose to examine deep learning for estimating both the dielectric and conductivity profiles of different types of complex walls. We study three different DNN architectures -a fully connected neural network (FC-NN), a convolutional neural network (CNN), and a generative adversarial network (GAN) for estimating the electrical characteristics of \emph{inhomogeneous lossy} walls. 
FC-NN is a multi-layered artificial neural network that models highly non-linear input-output relationships and has been used extensively for diverse pattern recognition problems. CNN introduces additional convolution and pooling layers to a fully connected layer and has been found to be extremely effective for image analysis problems. In contrast to FC-NN and CNN, GAN consists of two neural networks - a generator and a critic - that are configured in a game-theoretic framework. Each of the networks could be either ANN or CNN \cite{goodfellow2020generative}. Despite the increased computational complexity involved with training two networks, GANs are popular because they are more robust to smaller and less diverse training datasets compared to other algorithms with single networks.

For experimental validation, we generate both simulated electric field data from finite difference time domain techniques (FDTD), as well as measured frequency domain electric field data, gathered using a through-wall radar hardware setup. The scattered electric field data are presented to the optimized neural networks to solve the inverse scattering problem. We demonstrate that we can correctly estimate the thickness of the walls, dielectric profile, and conductivity profile up to 95\%, 96\%, and 90\% accuracy, respectively, showcasing the versatility of the DNNs-based algorithms for EIS problems. Most remarkably, we show that the DNNs trained on simulated data are capable of estimating the characteristics of real-world walls from their measured electric field data. Further, we show that the GAN-based network is the most robust with limited training data. 
 
 Our paper is organized in the following manner. In the next section, we describe the three different DNN approaches for estimating wall properties from the scattered electric field. In Section.\ref{sec:Sim}, we describe the experimental setup for generating simulated data that are subsequently used for training the DNNs and later validating the DNNs using test simulated data. Once validated, the DNNs are tested on measurement data from actual walls in Section.\ref{sec:Measurement}. This is followed by the conclusion and a discussion of the results in Section.\ref{sec:Conclusion}.
\section{Proposed Methodology}
\label{sec:Methodology}
\begin{figure}[htbp]
\centering
\includegraphics[scale=0.5]{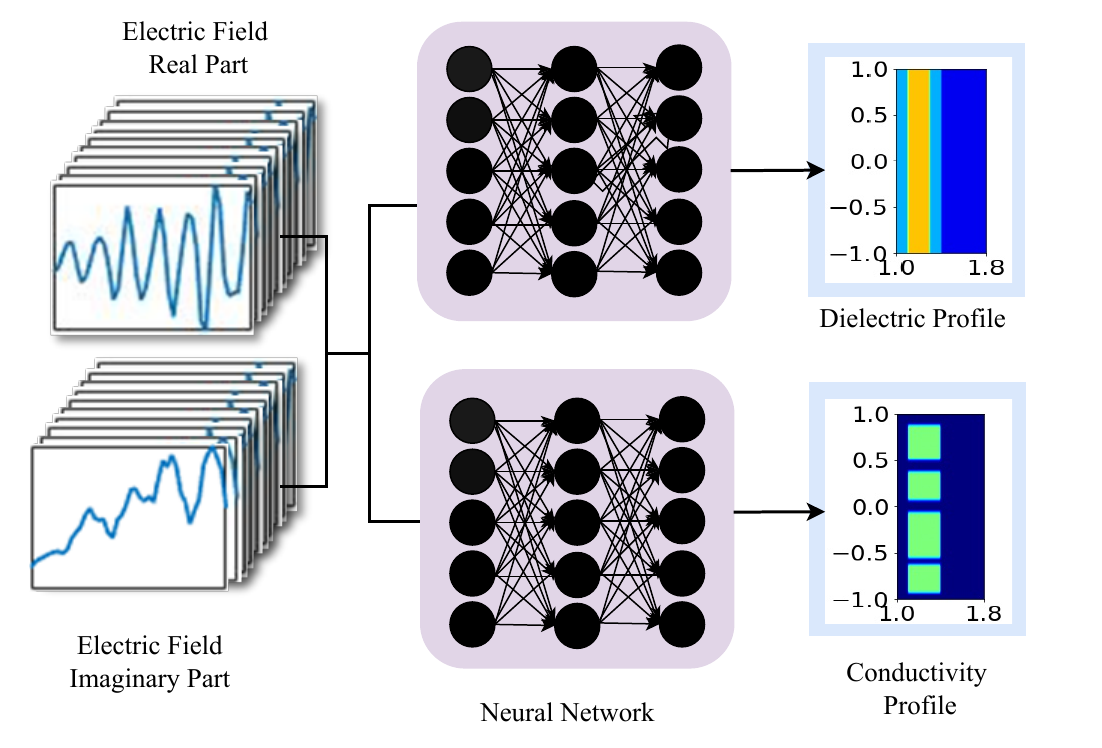}
\caption{System diagram depicting electromagnetic inversion operation using FC-NN/CNN for obtaining dielectric and conductivity profiles of walls.}
\label{fig:Block_diagram}
\end{figure}

\begin{figure}[htbp]
\centering
\includegraphics[scale=0.4]{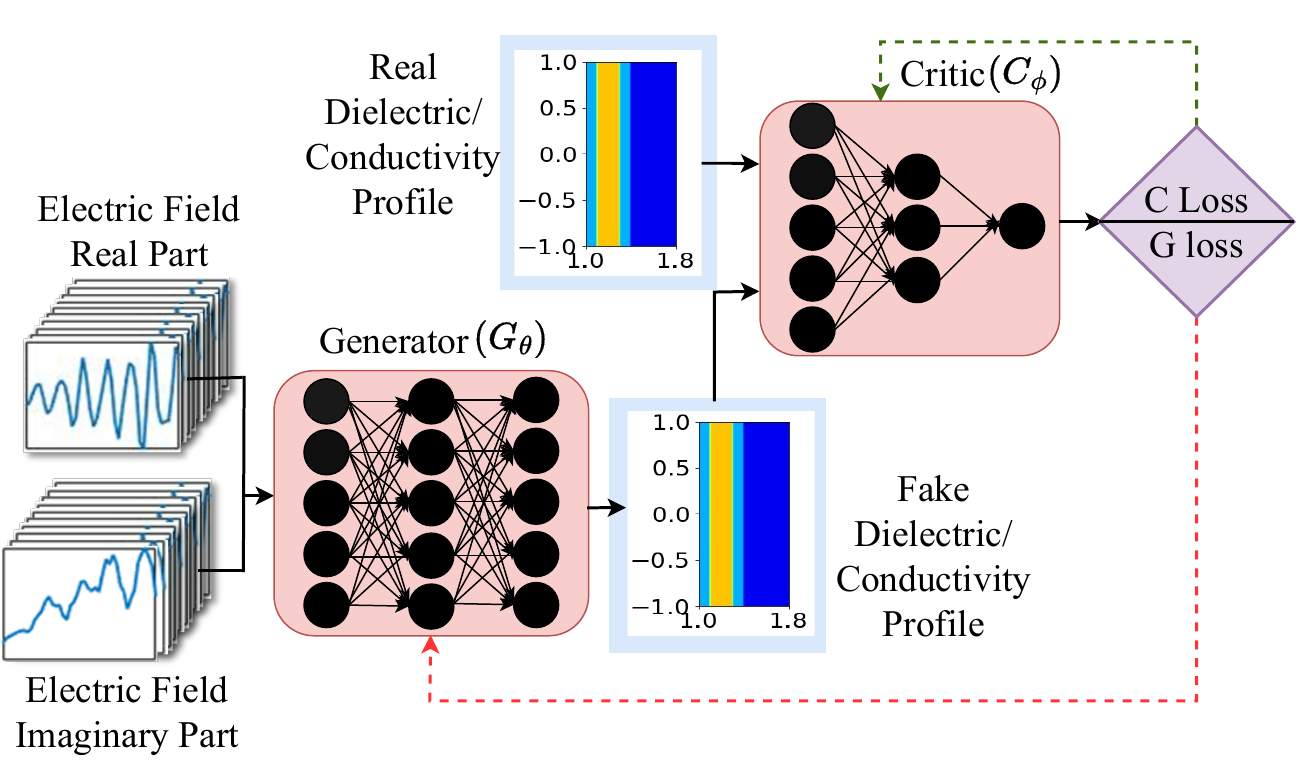}
\caption{System diagram depicting electromagnetic inversion operation using GAN for obtaining dielectric and conductivity profiles of walls.}
\label{fig:GAN_training}
\end{figure}

The objective of our work is to estimate the electrical characteristics of complex walls based on their scattered electric field. 
\begin{figure}[htbp]
\centering
\includegraphics[scale=0.55]{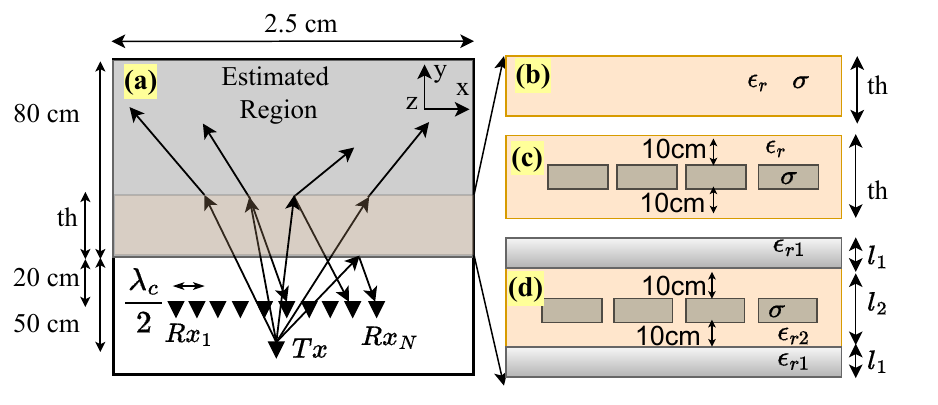}
\caption{(a) FDTD simulation setup for single transmitter and multiple element receiver array before (b) wall type-1: single layer homogeneous lossy dielectric wall, (c) wall type-2: single layer dielectric wall with periodic lossy regions, and (d) wall type-3 multiple layered dielectric walls with periodic lossy regions.}
\label{fig:simulation_setup}
\end{figure}
We assume a two-dimensional (2D) Cartesian problem space to minimize the computational complexity of the problem since most of the walls show homogeneity along their height, as shown in Fig.\ref{fig:simulation_setup}. Broadband vertically polarized electric field ($E_z^i$) is incident upon the wall from the transmitter, $Tx$, and the frequency domain scattered field ($E_z^{s},n=1:N$) from the wall is recorded at multiple receiver locations ($Rx_1 \cdots Rx_N$) on the same side of the wall as the transmitter. Measurements are made when there are no targets on the far side of the wall (opposite side to the transmitter). Therefore, the scattered field is only a function of the dielectric profile $\epsilon_r(x,y)$ and conductivity profile $\sigma(x,y)$, which are assumed to be fairly constant across the bandwidth. We use DNNs to perform the inversion operation from the scattered electric field to estimate $\epsilon_r(x,y)$ and $\sigma(x,y)$ of the region of interest beyond the front face of the wall. The region of interest is discretized along the two dimensions to form grids. The basic block diagram of the architecture is depicted in Fig.\ref{fig:Block_diagram}, where we consider two neural networks.
During training, $M$ samples of the frequency domain scattered electric field are provided as input to the two networks. Each sample is a column vector of length $P$ consisting of the real part of the electric field concatenated with the imaginary part of the field sampled at a frequency resolution of $\Delta f$. The output of the first network is a $M \times Q$ matrix consisting of $M$ samples of vectorized $\epsilon_r(x,y)$ corresponding to the spatial grid of the region of interest. The output of the second network is a $M \times Q$ matrix consisting of multiple samples of the vectorized $\sigma(x,y)$ at the same spatial resolution as the dielectric profile. 
We consider three different configurations of neural networks. The first is a fully connected multiple-layered artificial neural network (FC-NN); the second is a convolutional neural network (CNN) that consists of additional convolutional and spooling layers along with the fully connected layers; the third is a generative adversarial network which consists of two networks known as a critic ($C_{\phi}$) with weights $\phi$ and generator ($G_{\theta}$) with weights $\theta$ as shown in the Fig.\ref{fig:GAN_training}. During the training of FC-NN and CNN, the electric field data, and the corresponding dielectric and conductivity profiles, are provided as input and output, respectively, to the neural network. During validation/testing, we provide the test sample electric field and estimate the dielectric and conductivity profiles from the trained neural networks.
In the case of GAN training, we have two neural networks each, namely the generator and the critic, as shown in Fig.\ref{fig:GAN_training}, for estimating the dielectric and conductivity profiles. We provide the scattered electric field from the receiver positions as input to the generator. The outputs of the generator are synthetic wall profiles (dielectric/conductivity) which are provided as the input to the critic. The goal of the generator is to map the wall profile to a scattered electric field while implicitly training the distribution of the fake profile samples to closely resemble the distribution of the real profile samples. Together, both networks of the GAN work in an adversarial manner where the weights of the generator and critic are optimized based on a value function $V(G_{\theta}, C_{\phi})$ as given in:
\begin{multline}
\label{eqn:objectivefn}
\min_{G_{\theta}} \max_{C_{\phi}} V(G_{\theta},C_{\phi})=\min_{G_{\theta}} \max_{C_{\phi}} log(C_{\phi}(\epsilon_{r})) \\ +\log(1-C_{\phi}(G_{\theta}(E_z^{s})))
\end{multline}
The weights and biases of the generator are updated by keeping the critic constant and vice versa. The situation is a zero-sum game since each network cannot control the other network's parameters, but the loss function is dependent on both. 
During validation/test, we provide a test scattered electric field to the generator to estimate the wall profile.

On obtaining the vectorized dielectric or conductivity profiles from the different variants of the neural networks, we reshape the $\epsilon_r(x,y)$, and $\sigma(x,y)$ profile into a 2D profile as depicted in Fig.\ref{fig:Block_diagram}.
\section{Simulation Setup}
\label{sec:Sim}
In this section, we explain the simulation methods using the finite-difference time-domain (FDTD) technique to obtain the network training data consisting of scattered electric field data from complex wall scenarios \cite{ram2009simulation,ram2010simulation}. We consider a 2D simulation space spanning 2.5m along $x$ and $y$ directions as shown in Fig.\ref{fig:simulation_setup}.
We consider an infinitely long line source excitation corresponding to the transmitter at $(0m,0.5m)$, which gives rise to transverse magnetic wave propagation in the 2D space. The source consists of a Gaussian pulse excitation with a center frequency of $f_c=2.4GHz$ and a bandwidth of 2GHz, as shown in
\begin{align}
\label{eq:source_excitation}
S(t) = \frac{1}{\sqrt{2\pi \sigma_t^2}}10^{\frac{-(t-\mu_t )^2}{2\sigma_t^2}}\sin(2\pi f_{c}t)
\end{align}
where $\sigma_{t}$ and $\mu_t$ are 0.13ns and 2ns. The duration of the simulation is set as 21.5ns with a time resolution of 0.02ns to ensure Courant stability conditions \cite{zheng1999finite}. 
The simulation space is bounded by a perfectly matched layer of $2\lambda_c$ thickness where $\lambda_c$ corresponds to the wavelength. The entire 2D space is discretized into uniform grids that are one-tenth the wavelength resolution. The electric field propagates from the source and impinges upon the wall. Some of the energy propagates through the wall while the remaining is scattered. The frequency-domain electric field is recorded along ten positions equispaced from -0.28m to +0.28m along $x$, at a standoff distance of 0.2m behind the front face of the wall. The FDTD simulations are then repeated for free-space conditions with the same source excitation, and the electric field is recorded at the same ten receiver positions. Then the electric field of the free space scenarios is subtracted from that of the wall scenarios to remove the direct coupling between the transmitter and the receivers.
 
We consider three types of walls that are assumed to be uniform along $z$. All of these wall types resemble actual walls encountered in real-world scenarios. For each wall type, we consider multiple instances with distinct electrical characteristics as specified in Table.\ref{table:total_cases}. The first wall type (henceforth referred to as \emph{wall type-1}) is a lossy homogeneous dielectric wall whose relative permittivity ($\epsilon_r$) is varied between 4 and 8  and whose conductivity varies from $10^{-4}$ to $10^{-2}$ S/m to model the properties of common interior wall materials such as wood, brick, cement, and mud \cite{balanis2012advanced} while its thickness ($th$) is varied uniformly from 10cm to 50cm. The second wall type (\emph{wall type-2}) is a dielectric wall with multiple lossy regions. In this case, the dielectric constant of the wall material $\epsilon_r$ is varied uniformly from 4 to 8, and the thickness, $th$, is varied from 20cm to 50cm. Further, we have considered the four lossy regions in the wall while maintaining a constant edge thickness of the region to 10cm, as shown in Fig.\ref{fig:simulation_setup}b. The last wall type (\emph{wall type-3}) is an inhomogeneous wall with three dielectric layers aligned in the $y$ direction corresponding to a dielectric wall with insulation materials/facades on either side. The inner layer of thickness $l_2$ corresponds to a higher dielectric constant ($\epsilon_{r_2}$) than that of the two outer layers ($\epsilon_{r_1}$) of thickness $l_1$. We have considered periodic lossy regions inside the inner layer of the wall while maintaining an edge thickness of 10cm, as shown in Fig.\ref{fig:simulation_setup}d.
\begin{table}[!htbp]
\centering
\caption{Dielectric constant and thicknesses of walls}
\label{table:total_cases}
\begin{tabular}{*5c}
\hline 
\noalign{\vskip 1pt}
    {Wall type}&{Parameter}&{Values}&{cases}&{total cases}\\
\hline 
\noalign{\vskip 1pt}
     {Wall type 1}&{$\epsilon_{r}$}&{4-8}&{21}&{}\\
     {}&{$th$}&   {10-50cm}& { 5}& {}\\
     {}&{$\sigma$ (S/m)}&   {$10^{-4}$-$10^{-2}$}& {3}& {315}\\\hline 
     {Wall type 2}&  {$\epsilon_{r}$}& {4-8}&  {21}&{}\\
        {}& {$th$}&  {20-50cm}& {4}&{}\\
      {}& {$\sigma$ (S/m)}&{$10^{-4}$-$10^{-2}$}&{3}&{252}\\\hline
{Wall type 3}&{$\epsilon_{r_2}$}&  {4-8}&{5}&{}\\
  {}& {$\epsilon_{r_1}$}&  {2-3}& {2}& {}\\
  {}&            {$l_2$}&  {20-40cm}& {5}& {} \\
  {}&            {$l_1$}&  {5-10cm}& {2}&{}\\
  {}&{$\sigma$ (S/m)}&   {$10^{-4}$-$10^{-2}$}& {3}& {300}\\\hline 
\end{tabular}
\end{table}
In total, we consider 867 distinct wall cases. We perform Fourier transform on the time-domain electric field recorded at the ten receiver positions and then extract the complex components corresponding to the band of frequencies between 1.4 to 3.4GHz at a resolution of $\Delta f = 46.5$MHz.  The real parts of the fields are concatenated with the imaginary parts of the field to form a $[880 \times 1]$ vector, which is provided as input to the FC-NN, CNN, and GAN, respectively. The dielectric and conductivity profiles correspond to a spatial extent of the 2D Cartesian space spanning between -1m to +1m along $x$ and 1m to 1.8m along $y$ with a spatial resolution of 0.01m. Note that the spatial resolution for the DNNs corresponds to $32 \times 32$ pixel size of the dielectric profile which is lower than the spatial resolution of the FDTD grid space. A high-resolution profile that would match the FDTD space would result in much higher computational costs (both time and memory) of the  neural networks and hence is not adopted. The spatial extent of the mapped dielectric profile or conductivity profile in the $y$ direction of 0.8m is greater than the thickness of most real-world walls and chosen so as to enable the estimation of the actual wall thickness. 
\section{Network Details}
\label{sec:network_details}
Due to space constraints, we do not provide details of the hyper-parameter tuning of each of the networks and instead present the details of the optimized networks. 

\noindent\textbf{FC-NN configuration:} In the first method, we configure the FC-NN with two hidden layers of 256 and 512 nodes followed by an output layer with 1024 nodes. \emph{ReLU} and \emph{sigmoid} are the activation functions of the hidden layers and output layer, respectively. The learning rate is set at $2 \times 10^{-4}$ with Adam optimizer. The number of epochs is $100$ with a batch size of $32$, and binary cross entropy is the loss function.

\noindent\textbf{CNN configuration:} We configure the CNN network with two one-dimensional convolutional layers, where the first layer has $64$ filters with a kernel size of $3$, and the second layer has 64 filters with a kernel size of $3$, followed by the dense layer with $512$ nodes and an output layer with $1024$ nodes. The hidden layers have \emph{ReLU} activation function, whereas the output layer has a \emph{sigmoid} activation function. The network is configured with a learning rate of $1 \times 10^{-4}$ and Adam optimizer. The number of epochs is $100$ with a batch size of $32$ and binary cross entropy as a loss function.

\begin{figure}[htbp]
\centering
\includegraphics[scale=0.4]{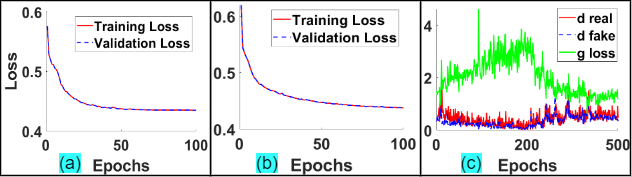}
\caption{Training and validation loss of (a) FC-NN, (b) CNN, and (c) generator and critic loss of GAN network.}
\label{fig:Loss_graph}
\end{figure}

\noindent\textbf{GAN configuration:} 
The generator of the GAN is configured with two convolutional layers - the first layer has 64 filters with a kernel size of $3$, and the second layer with 64 filters with a kernel size of $3$ - followed by a dense layer with $512$ nodes and an output layer with $1024$ nodes. The hidden layers have \emph{ReLU} activation function, whereas \emph{tanh} is the output layer activation function. The critic network consists of two hidden layers, each with a one-dimensional convolutional layer. Each hidden layer consists of 64 filters with a kernel size of 3 with \emph{ReLU} activation function. The output layer has one node with a $sigmoid$ function. 
We provide ground truth electric profiles of size $1024 \times 1$ as input to the critic. The networks are configured with a 0.0002 learning rate, Adam optimizer, where the number of the epochs is 500 with a batch size of 32, and binary cross entropy is the loss function. The training losses of all three architectures are presented in Fig.\ref{fig:Loss_graph}. 
For both FC-NN and CNN, as shown in Fig.\ref{fig:Loss_graph}a and Fig.\ref{fig:Loss_graph}b, we observe from the training and validation losses that the model is trained correctly without either overfitting or underfitting. In Fig.\ref{fig:Loss_graph}c, we represent the training loss of the generator and the critic's real and fake losses for 500 epochs. We observe that the generator loss (g loss) is converging, and the critic's loss for real (c real) and fake (c fake) samples lies around 0.5, which indicates that effective training has occurred. \cite{radford2015unsupervised}.
\section{Simulation Results}
\label{sec:sim_results}
In this section, we validate the effectiveness of the proposed DNN algorithms by comparing the estimated wall profiles from the neural networks with ground truth profiles. We divide the data into training and validation sets and check the performance of the networks for different percentages of training and validation volumes.
\subsubsection{Comparison of DNN methods}
In this section, we compare the inversion operation performance metrics for FC-NN, CNN, and GAN both visually and quantitatively. 
\begin{figure}[htbp]
\centering
\includegraphics[scale=0.27]{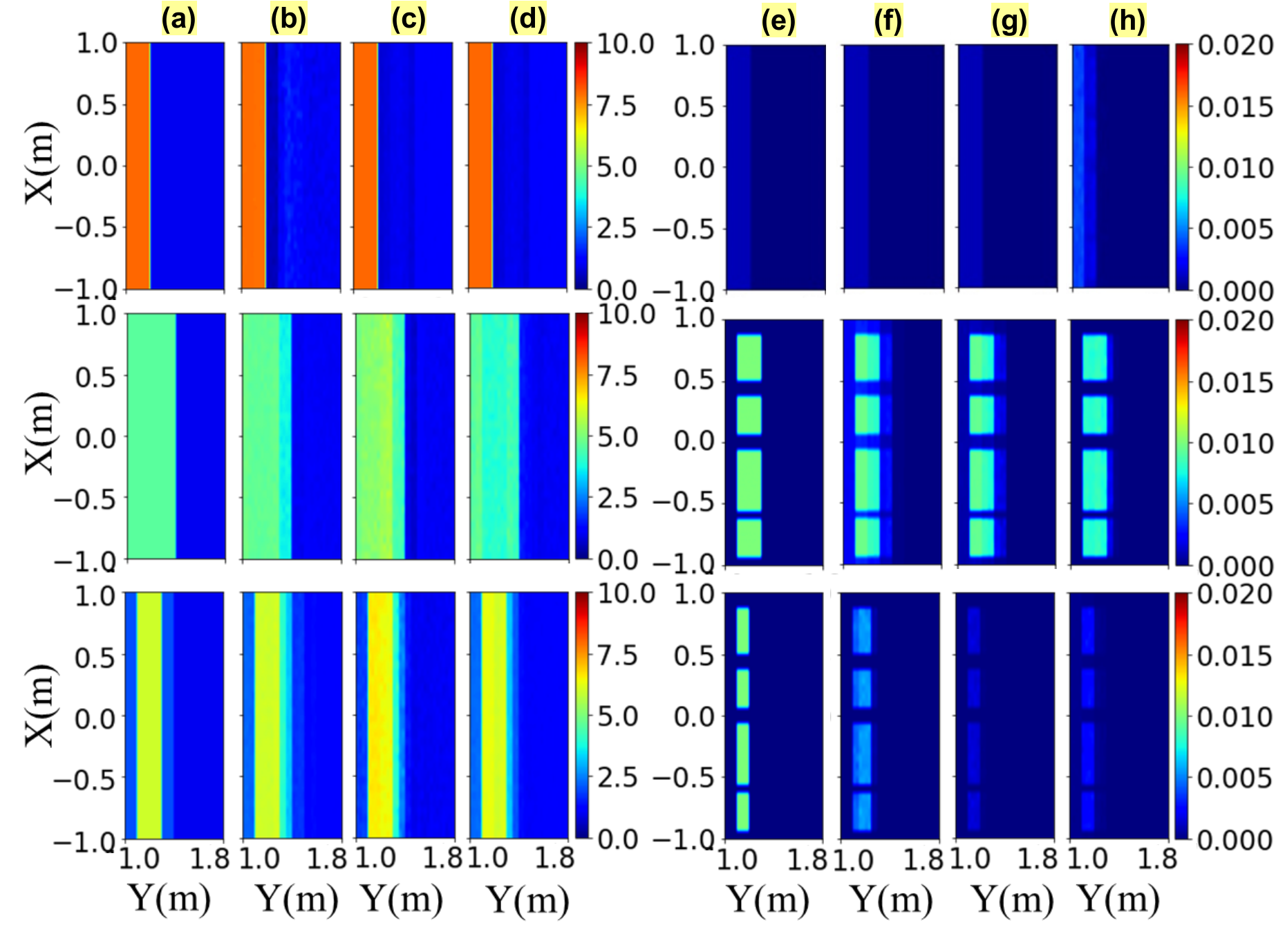}
\vspace{-2mm}
\caption{Estimation of dielectric profiles (a)-(d) and conductivity profiles (e)-(h) of (i) wall type-1 in top row, (ii) wall type-2 in second row, and (iii) wall type-3 in bottom row. Profiles shown in (a),(e) correspond to ground truth; (b) and (f) correspond to FC-NN, (c) and (g) correspond to CNN, and (d) and (h) correspond to GAN.}
\label{fig:estimated_profile}
\end{figure}
First, we present the qualitative results for one test case for each wall type. In Fig.\ref{fig:estimated_profile}, we present the dielectric profiles in the left set (columns (a) - (d)) and the conductivity profiles in the right set ((e) - (h)). The results from the rows correspond to  wall type-1 (top), wall type-2 (middle), and wall type-3 (bottom). The first column in each set ((a) and (e)) correspond to the ground truth profiles, while the remaining three columns in each set correspond to results generated from FC-NN, CNN, and GAN. In this case of wall type-1, the homogeneous wall is 20cm thick with a dielectric constant of 8 and conductivity of 0.001S/m. We observe from the results of FC-NN, CNN, and GAN that both the electrical characteristics and the thickness of the wall are reconstructed accurately. 

In the case of wall type-2, the wall is 50cm thick with a dielectric constant of 5.5 with regions of high conductivity of 0.01S/m, as shown in the ground truth profile. Again we observe from the reconstructed profiles for both the dielectric constant and the conductivity that the wall thickness is estimated correctly from the dielectric profile by all three DNN algorithms. The FC-NN's estimation of the conductivity profile is slightly more erroneous compared to the CNN and GAN. However, we are still able to observe the distinct lossy regions in the reconstructed profiles. Finally, we consider the last wall type of 50cm thickness with multiple dielectric layers ($(\epsilon_{r_1}=2,l_1=10cm,\epsilon_{r_2}=6,l_2=30cm)$.) and periodic lossy regions of 0.01S/m conductivity. We observe from the reconstructed profiles that the dielectric constants are estimated fairly accurately. However, there is some error in accurately reconstructing the correct thicknesses of the different layers. The reconstructed conductivity profiles by the three DNN algorithms show that the distinct lossy regions are correctly reconstructed. However, the estimates of the actual value of the conductivity show a greater error. The FC-NN estimates the conductivity to be slightly lower than the actual value. The CNN shows the maximum error (very low conductivity), and the GAN shows an intermediate error. 
The results indicate that the algorithms demonstrate a greater capability of reconstructing dielectric profiles accurately when compared to conductivity profiles. Second, greater inhomogeneity increases the overall reconstruction error.

For the quantitative comparison, the normalized mean square error (NMSE) is computed between the ground truth ($\sigma$) and estimated conductivity profiles ($\hat{\sigma}$) using
\begin{equation}
    \label{eqn: NMSE}
   NMSE = \frac{\left \| \sigma-\hat{\sigma}\right \|_{2}^{2}}{\left \|\sigma  \right \|_{2}^{2}}.
\end{equation}
The NMSE is computed across all test cases of data comprising all three types of walls. Similarly, we calculated NMSE between the ground truth ($\epsilon_r$) and estimated dielectric profiles ($\hat{\epsilon_r}$). The quantitative comparison of the techniques for all the test cases is summarized in Table.\ref{table:Comparison_methods}.
\begin{table}[!htbp]
\centering
\caption{Comparison of DNN methods for reconstructing electrical characteristics of complex walls.}
\label{table:Comparison_methods}
{\begin{tabular}{*5c}
\hline 
\noalign{\vskip 1pt}
    {Method}&{Electrical}&{{Avg. NMSE}}&{Training}&{Testing}\\
    {}&{Characteristics}&{}&{Time(min)}&{Time(s/min)}\\
\hline 
\noalign{\vskip 1pt}
     {FC-NN}&{Dielectric}&{0.05}&{30-45}&{0.01}\\
     {CNN}&{Dielectric}&{0.04}&{30-45}&{0.01}\\
     {GAN}&{Dielectric}&{0.04}&{120-150}&{0.01}\\ 
     {FC-NN}&{conductivity}&{0.15}&{30-45}&{0.01}\\
     {CNN}&{conductivity}&{0.1}&{30-45}&{0.01}\\
     {GAN}&{conductivity}&{0.1}&{120-150}&{0.01}\\
     \hline 
\end{tabular}}
\end{table}
Here, we observe that all methods demonstrate overall low NMSE and provide a promising indication of their capability to reconstruct wall profiles. However, we see that as we decrease the number of training sets for each algorithm, the performance goes down. However, GAN performs better in comparison to the other two methods for less volume of training data. In other words, the use of two adversarial networks for estimating each electrical profile results in greater robustness to limited training data. 
Similarly, we observe that the training time required to train the FC-NN and CNN is comparatively lesser than the GAN since the latter has two networks per profile to train. 
\begin{table}[!htbp]
\centering
\caption{Comparison of algorithms for different proportions of training to test data sets for reconstructing dielectric and conductivity profiles of complex walls.} 
\label{table:Comparison_train}
\begin{tabular}{*6c}
\hline 
\noalign{\vskip 1pt}
    {$Train\%$}&{$Test/Validation\%$}&{Profile}&{FC-NN}&{CNN}&{GAN}\\
\hline 
\noalign{\vskip 1pt}
     {$90\%$}&{$5\%$}&{$\hat{\epsilon_r}$}&{0.05}&{0.04}&{0.04}\\
     {$80\%$}&{$10\%$}&{$\hat{\epsilon_r}$}&{0.09}&{0.05}&{0.04}\\
     {$70\%$}&{$15\%$}&{$\hat{\epsilon_r}$}&{0.1}&{0.1}2&{0.1}\\
     {$90\%$}&{$5\%$}&{$\hat{\sigma}$(S/m)}&{0.15}&{0.12}&{0.15}\\
     {$80\%$}&{$15\%$}&{$\hat{\sigma}$(S/m)}&{0.26}&{0.26}&{0.27}\\
    {$70\%$}&{$10\%$}&{$\hat{\sigma}$(S/m)}&{0.34}&{0.32}&{0.26}\\
     \hline
\end{tabular}
\end{table}
Further, we observe that the estimated conductivity profile has a higher NMSE than the dielectric profile, possibly because the loss tangent ($\frac{\sigma}{\omega \epsilon_0}$) is a function of frequency even when the conductivity is fixed. Further, in our problem, we have considered more complex geometries for loss. The training and test codes are run with Keras 2.7 and trained and tested on an Intel Core i7-10510U processor running at 1.80 GHz. 
\section{Measurement Results}
\label{sec:Measurement}
In this section, we estimate the dielectric and conductivity profiles of actual walls using the proposed neural network that was trained using the simulated electric field data from the previous section. We feed measurements of the scattered electric field from real walls into the neural network, and the neural network returns an estimate of the wall's profile. It should be noted that the exact ground truth dielectric profile of the walls cannot be obtained. Therefore, we obtain a rough estimation of the wall's profile based on the material used in its construction.
\subsection{Measurement Setup}
\begin{figure}[htbp]
\centering
\includegraphics[scale=0.32]{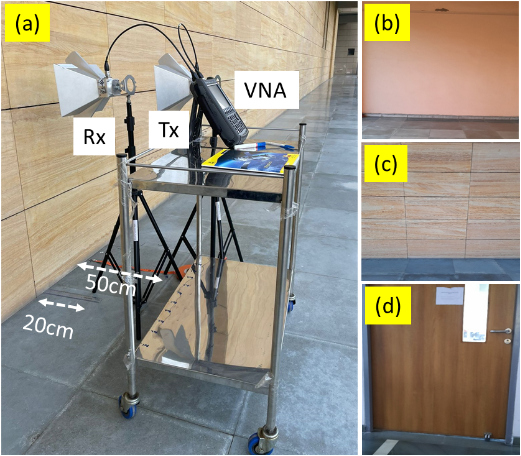}
\caption{(a) Experimental radar setup for gathering scattered electric field data before different types of wall: (b) 30cm thick exterior brick and concrete wall, (c) 40cm thick exterior brick wall with a thin facade of ceramic tiles, and (d) 3.5cm thick door.}
\label{fig:Experimental_setup}
\end{figure}

The measuring setup used to collect the broadband scattered electric field from the actual walls is depicted in Fig.\ref{fig:Experimental_setup}a. It consists of a two-port Field Fox N9926A vector network analyzer and two HF907 broadband horn antennas configured to measure the $S_{21}$ scattering parameter at center frequency 2.4 GHz for 2GHz bandwidth. The transmitted power is set at $3dBm$, and the sampling frequency of the measurement is 377Hz. The experimental setup is replicated to be as similar as possible to the simulation setup. We gather the scattered electric fields through a receiver placed at a standoff distance of 20cm and a transmitter placed 50cm before the front face of the wall. The receiver is moved along a straight lateral line parallel to the wall, and measurements are recorded at ten positions at 6.25cm spacing.
The scattered signal from the wall is captured by the VNA, where the received signal is amplified and then phase-quadrature demodulated and digitized. Next, the complex VNA measurements are collected and downloaded to a laptop for further processing. We conducted the measurements for two exterior walls and one door, as shown in Fig.\ref{fig:Experimental_setup}. The exterior wall is a 30cm thick brick wall with concrete and paint on either side. The second exterior wall is a 40cm brick wall with concrete on one side and a very thin layer of ceramic tiles having a thickness of 10mm, forming a facade on the other side. We also considered a 3.5cm wooden door. The dielectric constant of brick ranges from 3.8 and 4.2, stones range from 6 and 8, concrete ranges from  7.7 and 9.6, and ceramic tiles in between 21 and 28 respectively \cite{stavrou2003review,grosvenor2009time}. The broadband $S_{21}$ measurements directly correspond to frequency domain measurements. 
At each receiver position, we collected 225 samples for each frequency point and averaged them. We repeated the measurements in outdoor free-space conditions. Then we calibrate the scattered data by subtracting the free space signal from the wall scattered data. In this way, we remove the effect of the direct coupling between the transmitter and the receiver. Also, since we collected the data at each receiver position independently (synthetic array), there is no mutual coupling in the measurement data.
\subsection{Results}
\begin{figure}[htbp]
\centering
\includegraphics[scale=0.27]{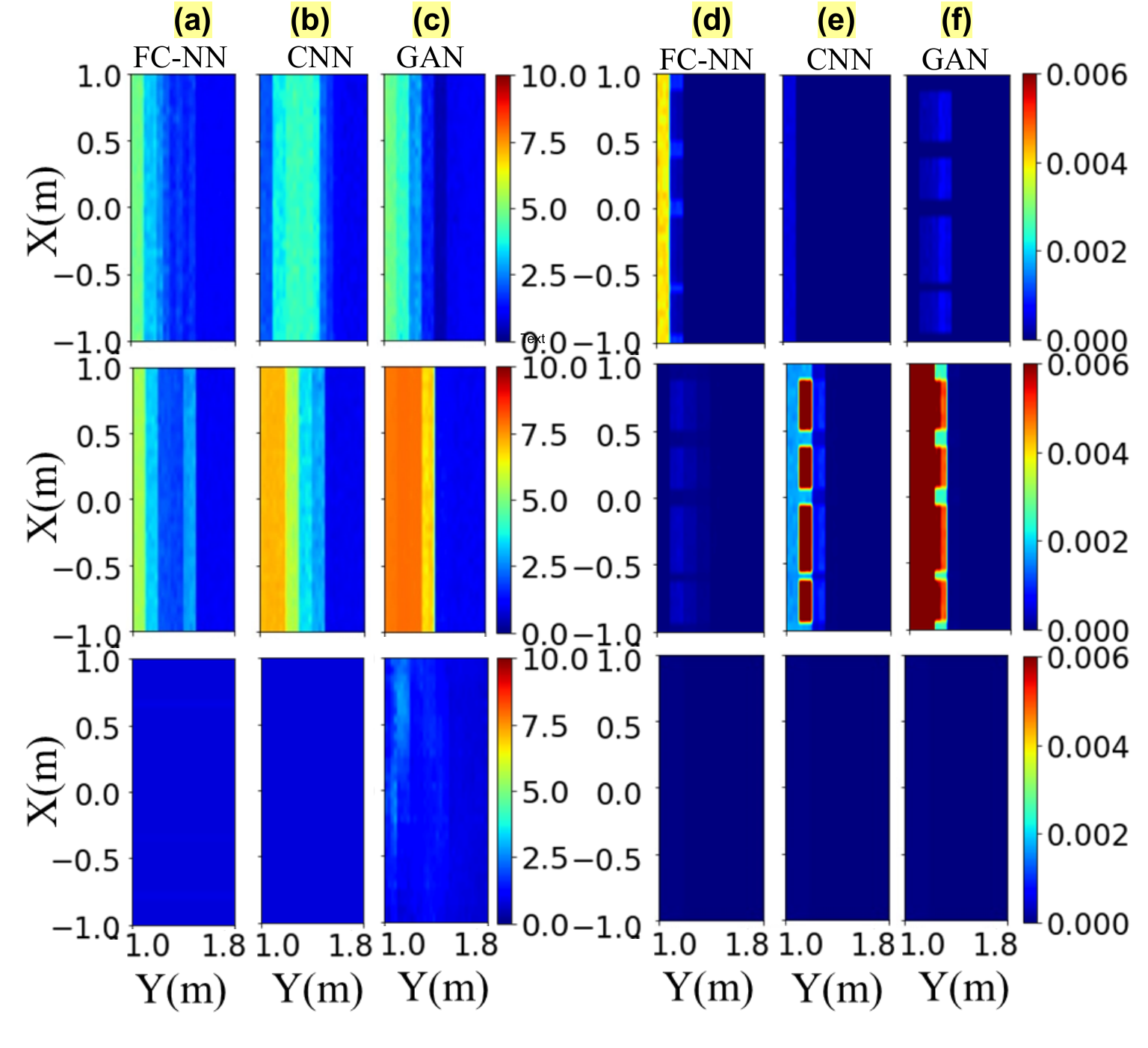}
\vspace{-1mm}
\caption{( Estimation of dielectric profiles (a)-(c) and conductivity profiles (d)-(f) of (i) 30cm thick exterior brick wall in the top row, (ii) 40cm thick exterior brick wall with a facade of ceramic tiles in the second row, and (iii) 3.5cm wooden door in the bottom row. Profiles shown in (a) and (d) correspond to FC-NN, (b) and (e) correspond to CNN, and (c) and (f) correspond to GAN.}
\label{fig:Exp_result}
\end{figure}

We have reconstructed the dielectric and conductivity profile from the neural networks trained with simulation data. The first column shows the reconstructed dielectric profiles from FC-NN, while the second column shows the results from CNN, and the third column shows results from GAN as shown in Fig.\ref{fig:Exp_result}. The first row depicts a 30cm thick exterior brick and concrete wall; the second row depicts a 40cm thick exterior brick wall with a thin facade of ceramic tiles, and the third row depicts a 3.5cm thick door. GAN estimated the thickness of the walls with an error of 5\%, whereas the thickness of walls with FC-NN and CNN were estimated with an error of  10\% and 12\%, respectively. This demonstrates that GAN can handle high dielectric values. The results in the top two rows show that the neural networks are able to reconstruct the dielectric profile. Here, we observe that the GAN estimated the thickness of the walls quite correctly in comparison to FC-NN and CNN. 
This corroborates our earlier findings with the simulation validation results, which showed that the GAN architecture was most robust to limited training data. Due to the significant difference between the measurement and simulation scenarios, the GAN trained with the simulation data was most effective on real-world measurement data. In the case of the 3.5cm wooden door, we are unable to reconstruct the dielectric and conductivity profiles as shown in Fig.\ref{fig:Exp_result}. This is because the door is mostly transparent to the electromagnetic field in the range of 1.4GHz to 2.4GHz, resulting in almost negligible scattered returns. A simple method to overcome this limitation is to increase the frequency of the source excitation. Fortunately, such walls do not cause significant distortions to through-wall radar signatures, and hence it is not imperative to remove the wall effects.
\section{Conclusions}
\label{sec:Conclusion}
We propose DNN methods configured by FC-NN, CNN, and GANs with frequency domain input scattered electric field data to solve to estimate the dielectric and conductivity profiles of complex inhomogeneous walls. We experimentally validate these techniques using both simulations and measurements. The results clearly demonstrate the effectiveness of the methods in reconstructing electric profiles of walls. Further, we demonstrate that while the single neural networks - FC-NN and CNN- require lesser training time and computational resources, the GANs configured with two networks are more effective with limited training data. We also demonstrated that the neural networks that are trained with simulated data are reasonably effective at determining the electrical characteristics of actual walls, provided these walls scatter significant energy.
\vspace{-2mm}
\bibliographystyle{ieeetran}
\bibliography{main}

\end{document}